\documentclass[multphys,vecphys]{svmult}
\usepackage{graphics,graphicx,dcolumn,bm,fleqn,epic,eepic,float}
\usepackage{amssymb,amsmath,multirow,rotate,color,float,times}
\usepackage{color}
\usepackage{makeidx}         
\usepackage{multicol}        
\usepackage[bottom]{footmisc}
\definecolor{red}{rgb}{1,0,0}
\definecolor{green}{rgb}{0,1,0}
\definecolor{blue}{rgb}{0,0,1}

\makeindex             

\begin{document}
\title*{Numerical Improvement of the Discrete Element Method applied
       to Shear of Granular Media}
\titlerunning{Modeling Shear of Irregular Particles}


\author{Andr\'es A.~Pe\~na\inst{1}\and
        Pedro G.~Lind\inst{1}$^,$\inst{2}\and
        Sean McNamara\inst{1}\and 
        Hans J.~Herrmann\inst{3}$^,$\inst{4}}
\institute{Institute for Computational Physics, 
           Universit\"at Stuttgart, Pfaffenwaldring 27, 
           D-70569 Stuttgart, Germany \and
           Centro de F\'{\i}sica Te\'orica e Computacional, 
           Av.~Prof.~Gama Pinto 2,
           1649-003 Lisbon, Portugal \and
%
%
           Departamento de F\'{\i}sica, Universidade Federal do Cear\'a,
           60451-970 Fortaleza, Cear\'a, Brazil \and
           Computational Physics, 
           IfB, HIF E12, ETH H\"onggerberg, CH-8093 Z\"urich,
           Switzerland}

\maketitle

\begin{quote}
\footnotesize
We present a detailed analysis of the bounds on the
integration step in Discrete Element Method (DEM) 
for simulating collisions and 
shearing of granular assemblies.
We show that, in the numerical scheme, the upper limit for the 
integration step, usually taken from the average time $t_c$ 
of one contact, is in fact not sufficiently small to guarantee 
numerical convergence of the system during relaxation.
In particular, we study in detail how the kinetic energy decays
during the relaxation stage and compute the correct upper limits for the 
integration step, which are significantly smaller than the
ones commonly used.
In addition, we introduce an alternative approach, based on
simple relations to compute the frictional forces,
that converges 
even for integration steps above the upper limit.
\normalsize
\end{quote}




\section{Introduction}

One of the standard approaches to model the dynamics of
granular media is to use the Discrete 
Element Method (DEM)~\cite{poeschel05,ciamarra05,dacruz05}, 
e.g.~to study shear~\cite{cundall89,thompson91,pena07,%
fernando06,mora99}. 
Some problems may arise due to the need to use
large integration steps to perform numerical 
simulations with reasonable computational effort,
without compromising the overall convergence of the numerical
scheme.
For slow shearing, the convergence of the numerical
schemes 
is particularly important
when studying for instance the occurrence of avalanches%
~\cite{mora99} and the emergence of 
ratcheting in cyclic loading~\cite{mcnamara07}.
Usually, one assumes an upper limit for the admissible integration
steps, based on empirical reasoning\cite{allen03}.

The aim of this paper is twofold.
First, we show that an integration step able to guarantee convergence 
of the numerical scheme, must in general be smaller than a specific
upper limit, significantly below the commonly accepted 
value~\cite{thompson91,allen03,luding94}.
This upper limit strongly depends on 
(i) the accuracy of the approach used to calculate frictional forces 
between particles,
(ii) on the corresponding duration of the contact and 
(iii) on the number of degrees of freedom.
Second, we address the specific case of slow shearing, for which
the above limit is too small to allow for reasonable
computation time.
To overcome this shortcoming we propose an alternative
approach that corrects the frictional contact forces, when large 
integration steps are taken.
In this way, we enable the use of considerably 
larger integration steps, 
assuring at the same time the convergence of the integration scheme.

We start in Sec.~\ref{sec:model} by presenting in some detail the 
Discrete Element Method~\cite{poeschel05,allen03,cundall79}.
Sections \ref{sec:sensitivity} and \ref{sec:geometrical}
describe respectively the dependence on the integration step and the 
improved algorithm.
Discussions and conclusions are given in 
Sec.~\ref{sec:conclusions}.
\begin{figure} [t]
\begin{center}
\includegraphics*[width=8.0cm,angle=0]{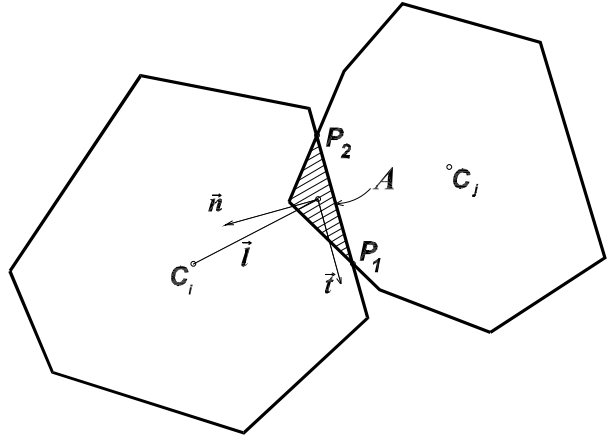}
\end{center}
\caption{\protect
    Illustration of two overlapping particles. 
    The overlap region $A$ between particles fully characterizes the
    contact force $\vec{F}^c$ (see text).} 
\label{fig1}
\end{figure}

\section{The model} 
\label{sec:model}

We consider a two-dimensional system of polygonal particles, 
each one having two linear and one rotational degree of 
freedom.
The evolution of the system is given by 
Newton's equations of motion, 
where the resulting forces and 
moments acting on each particle $i$ are given by the sum 
of all forces and momenta applied on that particle:
\begin{subequations}
\begin{eqnarray}
m_i \ddot{\vec{r}}_i &=& \sum_{c} \vec{F}^c_i +
     \sum_{c_b} \vec{F}^b_i  , \label{newton1} \\
  & & \cr
I_i \ddot{\vec{\theta}}_i &=& \sum_{c} \vec{l}^c_i \times
     \vec{F}^c_i  +  \sum_{c_b} \vec{l}^b_i \times
     \vec{F}^b_i  ,\label{newton2}
\end{eqnarray}
\label{eq:Newton}
\end{subequations}
where $m_i$ denotes the mass of particle $i$, $I_i$ its 
moment of inertia, and $\vec{l}$ the branch vector which 
connects the center of mass of the particle to the 
application point of the contact force $\vec{F}^c_i$ or 
boundary force $\vec{F}^b_i$.
The sum in $c$ is over all the particles in contact with polygon 
$i$, and the sum in $c_b$ is over all the vertices of 
polygon $i$ in contact with the boundary.
One integrates 
Eqs.~(\ref{eq:Newton}) for all particles $i=1,\dots,N$
and obtains the evolution of their
centers of mass $\vec{r}_i$ and rotation angles $\theta_i$.

Further, during loading, particles tend to deform each other.
This deformation of the particles is usually reproduced by 
letting them overlap~\cite{poeschel05,cundall79}, as illustrated
in Fig.~\ref{fig1}. The overlap between each pair of particles
is considered to fully characterize the contact.
Namely, the normal contact force is assumed to be proportional to
the overlap area~\cite{tillemans95} and its direction perpendicular to the plane of contact, which is defined by
the intersection between the boundaries of the two particles.


All the dynamics is deduced from the contact forces acting on
the particles.
The contact forces, $\vec{F}^c$, either between particles or 
with the boundary, are decomposed into their elastic and viscous 
contributions, $\vec{F}^e$ and $\vec{F}^v$ respectively,
yielding $\vec{F}^c = \vec{F}^e + \vec{F}^v$.

The viscous force is important for maintaining the numerical stability 
of the method and to take into account dissipation at the contact.
This force is calculated as~\cite{cundall79}
\begin{equation}
\vec{F}^v = - m_r \nu \vec{v}^c , 
\label{viscousforce}
\end{equation}
where $m_r=(1/m_i+1/m_j)^{-1}$ is the reduced mass of the two particles, 
$i$ and $j$, and $\nu$ is the damping coefficient.

The elastic part of the contact force is what will be carefully
studied, since it is what determines the accuracy of the integration 
scheme.
The term $\vec{F}^e$ is simply given by the sum of the normal and the 
tangential components, with respect to the contact plane, namely
\begin{equation}
\vec{F}^e = F^e_n \hat{n}^c + F^e_t \hat{t}^c ,
\label{elastContForc}
\end{equation}
where the normal component reads
\begin{equation}
F^e_n = -k_n A/l_c ,
\label{normalelastforc}
\end{equation}
with $k_n$ the normal stiffness, $A$ the overlap area and 
$l_c=r_i+r_j$ the characteristic length of the contact between
particles $i$ and $j$, with $r_i=\sqrt{A_i/2\pi}$ and $A_i$ the 
area of the particle $i$ (and similarly for particle $j$).

Using an extension of the Cundall-Stack 
spring~\cite{cundall79}, which considers the tangential force to 
be proportional to the elastic elongation $\xi$ of an imaginary 
spring at the contact, one defines the static frictional force between 
each pair of particles in contact, as
\begin{equation}
F^e_t = - k_t \xi ,
\label{tangelastforc}
\end{equation}
where $k_t$ is the tangential stiffness. 
This tangential force assumes each contact as being 
described by a damped oscillator with some frequency $\omega$ (see
below).

The elastic elongation $\xi$ in Eq.~(\ref{tangelastforc}) is updated as 
\begin{equation}
\xi(t+\Delta t) = \xi(t) + \vec{v}_t^c \Delta t  
\label{elasticelong}
\end{equation}
where 
$\Delta t$ is the time step of the DEM simulation, 
and $\vec{v}_t^c$ is the tangential component of the relative velocity 
\begin{equation}
\vec{v}^c = \vec{v}_i - \vec{v}_j + \vec{w}_i \times \vec{l}_i -
\vec{w}_j \times \vec{l}_j ,
 \label{eq:vc}
\end{equation}
at the contact point. 
Here, $\vec{v}_i$ and $\vec{v}_j$ are the linear velocities of the
centers of mass and $\vec{w}_i$ and $\vec{w}_j$ the 
angular velocities of the particles around the corresponding centers 
of mass.

The tangential elastic elongation
$\xi$ changes according to Eq.~(\ref{elasticelong}) 
whenever the 
condition $|F_e^t| < \mu F_e^n$ is satisfied, whereas, when the Coulomb 
limit condition $|F_e^t| = \mu F_e^n$ is reached, sliding is enforced by 
keeping constant the tangential force $F_e^t$ and assigning to $\xi$ its 
extreme values $\pm \mu k_n A/(k_t l_c)$.
This latter Coulomb condition corresponds to the regime where particles
behave inelastically, while the former inequality describes the forces
when the particles behave elastically.
Parameter $\mu$ is the inter-particle friction coefficient.

In DEM, one of the numerical integration schemes usually used
to solve the equations of motion above is Gear's 
predictor-corrector scheme~\cite{allen03}.
This scheme consist of three main stages, namely prediction, 
evaluation and correction.

In the prediction stage the linear and angular positions, velocities and 
higher-order time derivatives are updated by expansions in 
Taylor series using the current values of these 
quantities~\cite{allen03,rougier04}.
In particular, one extracts a predicted position $\vec{r}^p(t + \Delta t)$ 
and acceleration $\ddot{\vec{r}}^p (t + \Delta t)$ for the center of
mass of a given particle and the predicted angular
displacement $\vec{\theta}^p(t + \Delta t)$ and angular
acceleration $\ddot{\vec{\theta}}^p(t + \Delta t)$ of that particle around
its center of mass.


During the evaluation stage, one uses the predicted coordinates
to determine the contact force 
$\vec{F}^c_{t + \Delta t}$ at time $t + \Delta t$.
Since the method is not exact, there is a difference between
the acceleration $\ddot{\vec{r}}(t + \Delta t) = 
\vec{F}^c_{t + \Delta t} / m$ and the value obtained in the 
prediction stage, namely 
\begin{equation}
\Delta \ddot{\vec{r}} = \ddot{\vec{r}}(t + \Delta t) - 
                        \ddot{\vec{r}}^p(t + \Delta t).
\label{discrepancy}
\end{equation}

The difference $\Delta \ddot{\vec{r}}$ in Eq.~(\ref{discrepancy}) 
is then finally used in the corrector step to correct the predicted 
position and its time derivatives, using proper weights for each time 
derivative~\cite{allen03}, that depend upon the order of the 
algorithm and the differential equation being solved.
These corrected values are the ones used for the next integration
step $t+\Delta t$.
This same procedure is also 
applied to the rotation angles 
$\theta_i$ around the center of mass as well as to their time 
derivatives, yielding the correction $\Delta \ddot{\vec{\theta}}$.

In our simulations we integrate equations of the form 
$\ddot{\vec{r}} = f(\vec{r},\dot{\vec{r}})$, using a fifth order 
predictor-corrector algorithm that has a numerical error proportional 
to $(\Delta t)^6$ for each integration step~\cite{allen03}.
However, as will be seen in Sec.~\ref{sec:sensitivity}, 
$(\Delta t)^6$ is not the 
numerical error of the full integration scheme, since 
Eq.~(\ref{tangelastforc}), used to calculate the frictional force, 
is of order $(\Delta t)^2$.

For a certain value of normal contact stiffness $k_n$, almost any 
value for the normal damping coefficient $\nu_n$ might be selected. 
Their relation defines the restitution coefficient $\epsilon$ obtained 
experimentally for various materials~\cite{foerster94}. 
The restitution coefficient is given by the ratio between the relative 
velocities after and before a collision. 
In particular, the normal restitution coefficient $\epsilon_n$ can be
written as a function of $k_n$ and $\nu_n$~\cite{luding98}, namely
\begin{equation}
 \epsilon_n = \exp \ ( -\pi \eta / \omega)
            = \exp{\left ( -\frac{\pi}{\sqrt{4m_rk_n/\nu_n^2-1}}\right )}
\label{eq:Res_coeff}
\end{equation}
where $\omega = \sqrt{\omega_0^2-\eta^2}$ is the frequency of 
the damped oscillator, with $\omega_0=\sqrt{k_n/m_r}$ the frequency of 
the elastic oscillator, 
$m_r$ the reduced mass and $\eta=\nu_n/(2m_r)$ the effective viscosity.
The tangential component $\epsilon_t$ of the restitution coefficient
is defined similarly using $k_t$ and $\nu_t$ in Eq.~(\ref{eq:Res_coeff}).

Therefore, a suitable closed set of parameters for this model are
the ratios $k_t/k_n$ and $\epsilon_t/\epsilon_n$ (or $\nu_t/\nu_n$),
together with the normal stiffness $k_n$ and the interparticle friction 
$\mu$.

The entire algorithm above relies on a proper choice of the
integration step $\Delta t$, which should 
neither be too large to avoid divergence of the integration
nor too small avoiding unreasonably long computational time.
The determination of the optimal integration step varies from
case to case and there are two main criteria to estimate
an upper bound for admissible integration steps.
\begin{figure}[t]
\begin{center}
\includegraphics*[width=8.5cm,angle=0]{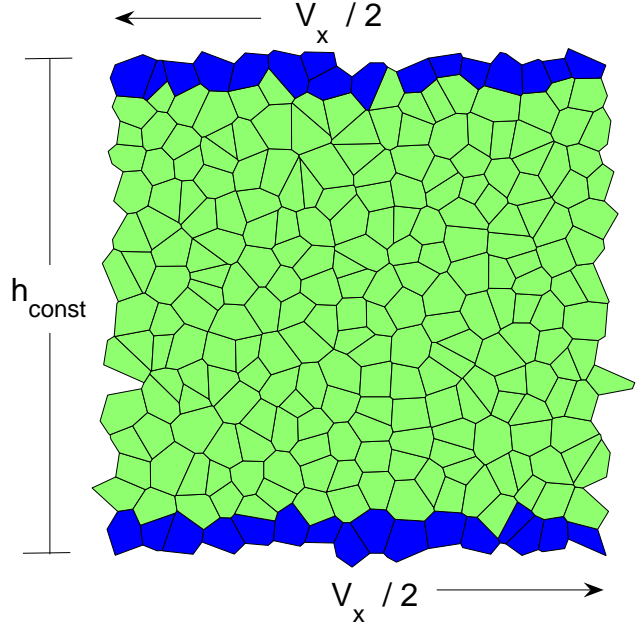}
\end{center}
\caption{\protect
     (Color online)
     Sketch of the system of $256$ particles (green particles)
     under shearing of top and bottom boundaries (blue particles). 
     Horizontally periodic boundary conditions are considered and 
     a constant low shear rate is chosen (see text).}
\label{fig2}
\end{figure}

The first criterion is to use the characteristic period of 
oscillation~\cite{allen03},
defined as
\begin{equation}
t_s = 2\pi \sqrt{\frac{\langle m\rangle}{k_n}} ,
\label{periodOsc}
\end{equation} 
where $\langle m\rangle$ is the smallest particle mass in
the system.
For a fifth order predictor-corrector integration scheme, it is
usually accepted that a safe integration step should be below
a threshold of $\Delta t < t_s/10$~\cite{allen03}.

The second criterion is to extract the threshold from local 
contact events~\cite{thompson91,luding94,luding98}, namely 
from the characteristic duration of a contact:
\begin{equation}
t_c = \frac{\pi}{\sqrt{\omega_0^2-\eta^2}} .
\label{tc}
\end{equation}
Typically $t_c\simeq t_s/2$, and therefore in such cases, one 
considers an admissible integration step as 
$\Delta t < t_c/5$~\cite{luding98,matuttis98}.

In the next section we will study in detail the integration for different 
values of the model parameters.

\section{The choice of the integration step}
\label{sec:sensitivity}

We simulate the relative motion of two
plates shearing against each other~\cite{fernando06,mora99,tillemans95}.
Considering a system of 256 particles as illustrated in Fig.~\ref{fig2},
where both top and bottom boundaries move in opposite directions
with a constant shear rate $\dot{\gamma}$.
The top and bottom layer of the sample have fixed boundary conditions, 
while horizontally we consider periodic boundary conditions.
The volumetric strain is suppressed, i.e.~the vertical position of 
the walls is fixed and there is no dilation. 
The particles of the fixed boundary are not allowed to rotate or move 
against each other. 
The shear rate is $\dot{\gamma}=1.25\cdot 10^{-5}  \hbox{s$^{-1}$}$,
the parameter values are $k_n=400 \hbox{\ N/m}$, $\epsilon_n=0.9875$,
and $\mu=0.5$. 
The relation $k_t/k_n$ is chosen such that $k_t < k_n$, similarly
to previous studies~\cite{cundall79,luding98,alonso04}, namely $k_t/k_n=1/3$.
Further, for simplicity we consider $\nu_t/\nu_n=k_t/k_n$, which 
when substituted in Eq.~(\ref{eq:Res_coeff}) yields 
$\epsilon_t/\epsilon_n=1.0053$.
\begin{figure}[t]
\begin{center}
\includegraphics*[width=9.5cm,angle=0]{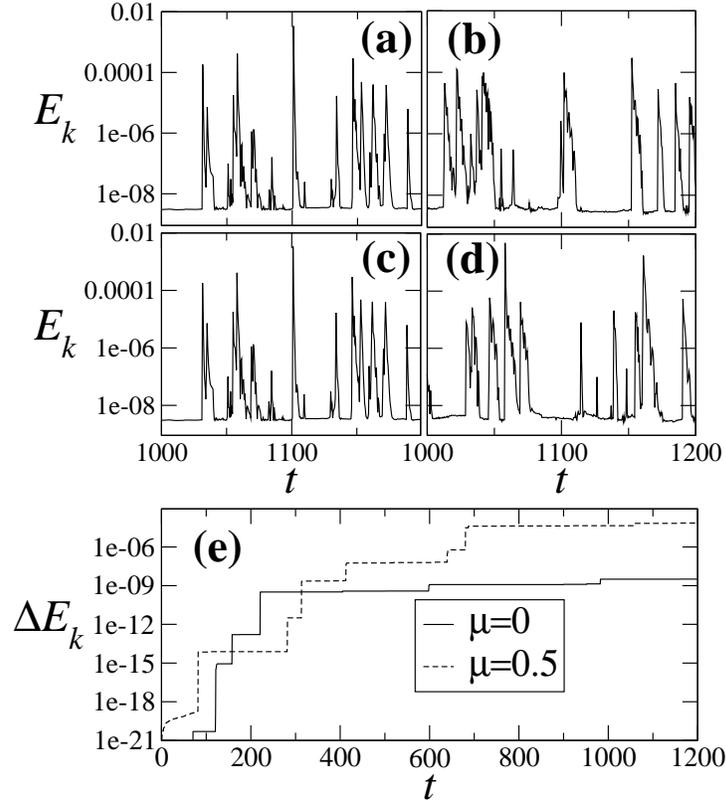}
\end{center}
\caption{\protect
     Dependence of the numerical scheme on the integration step
     $\Delta t$ and the friction coefficient $\mu$, by plotting the 
     kinetic energy $E_k$ as a function of time, for
     {\bf (a)} $\Delta t=10^{-3}\hbox{\ s}$ and $\mu=0$ (no friction),
     {\bf (b)} $\Delta t=10^{-3}\hbox{\ s}$ and $\mu=0.5$,
     {\bf (c)} $\Delta t=5\times 10^{-3}\hbox{\ s}$ and $\mu=0$, and
     {\bf (d)} $\Delta t=5\times 10^{-3}\hbox{\ s}$ and $\mu=0.5$.
     In {\bf (e)} we show the difference between the values of $E_k$
     obtained with the two values of $\Delta t$.
     Here, $k_n=400$ and the parametric relations 
     in Eq.~(\ref{periodOsc}) and (\ref{tc}) are used (see text).}
\label{fig3}
\end{figure}

By integrating such a system of particles using the scheme described
in the previous section, one can easily compute the kinetic
energy $E_k$ of a given particle $i$, 
\begin{equation}
E_k(i)=\frac{1}{2}\left (
       m_i\dot{\vec{r}}_i^2+I_i\vec{\omega}_i^2
       \right ) ,
\label{Ek}
\end{equation}
where velocity $\dot{\vec{r}}$ is computed from the predictor-corrector
algorithm, $I_i$ is the moment of inertia of the polygon
and $\vec{\omega}_i$    
is the angular velocity.

In Fig.~\ref{fig3} we show the evolution of the kinetic energy
for two different $\Delta t = 0.001 \hbox{\ s}$ and $0.005
\hbox{\ s}$.
As one sees, frictionless particles (Fig.~\ref{fig3}a and
\ref{fig3}c) have an $E_k$ that does not change for different integration 
steps, while when $\mu=0.5$ (Fig.~\ref{fig3}b and \ref{fig3}d) 
the evolution of $E_k$ strongly depends on $\Delta t$.
In Fig.~\ref{fig3}e we plot the cumulative difference $\Delta E_k$
between the values of $E_k$ taken for each integration step.
Here, one sees that in the absence of friction $\Delta E_k$ 
is significantly lower then when friction is present.

The two values of $\Delta t$ used in Fig.~\ref{fig3} can be
written as $\Delta t = 13/500 t_c$ and $13/2500 t_c$. Thus, we conclude
that the expected upper limit $\sim t_c/10$ is still too 
large to guarantee convergence of the integration scheme when friction 
is considered.

\begin{figure}[H]
\begin{center}
\includegraphics*[width=7.0cm,angle=0]{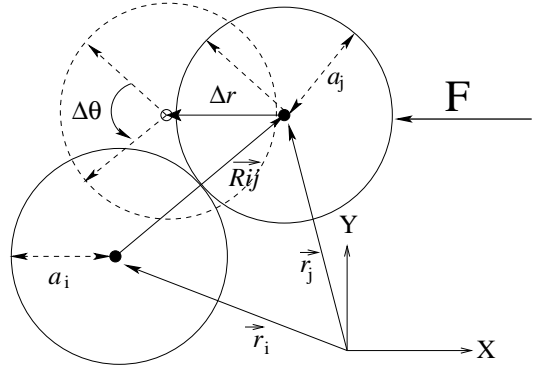}
\end{center}
\caption{\protect
     Sketch of the stress controlled test of two particles (discs).
     The particle located at $\vec{r}_i$ remains fixed, while the particle
     at $\vec{r}_j$ is initially touching particle $i$. The vector
     $\vec{R}_{ij}$ connecting the center of mass of particles $i$ and 
     $j$ is initially oriented $45^o$ with respect to the $x$-axis.
     After applying the constant force $\vec{F}$ to disc $j$,
     the system relaxes to a new position (dashed circumference). 
     Between its initial and final position particle $j$ undergoes 
     a displacement $\Delta r$ and a rotation $\Delta\theta$ 
     (see text).}
\label{fig4}
\end{figure}
\begin{figure}[htb]
\begin{center}
\includegraphics*[width=6.3cm,angle=0]{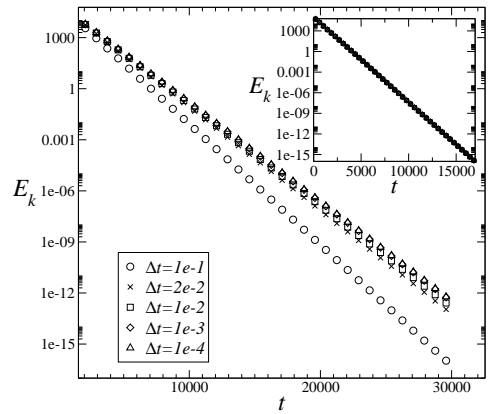}
\end{center}
\caption{\protect 
     The relaxation of the system of two discs sketched in
     Fig.~\ref{fig4}. Here we plot the kinetic energy $E_k$
     as a function of time $t$ (in units of $t_c$) for 
     different integration increments
     $\Delta t$ and using a stiffness $k_n=4\times 10^8\hbox{\ N/m}$ 
     and a friction coefficient $\mu=500$. 
     The large $\mu$ value is chosen so that the system remains in the
     elastic regime.
     As one sees, the 
     relaxation time $t_R$ converges to a constant value
     when $\Delta t$ is sufficiently small (see text).
     This discrepancy between the values of $t_R$ when different
     integration steps are used does not occur in the absence
     of friction ($\mu=0$), as illustrated in the inset.
     The slope of the straight lines is $-1/t_R$ (see 
     Eq.~(\ref{kinen_decay})).}
\label{fig5}
\end{figure}

Next we will perform a careful analysis to obtain a proper integration 
step as function of the parameters of our model.
For that, we consider the simple situation of two circular particles
and study the kinetic energy of one of them under external forcing,
as sketched in Fig.~\ref{fig4}.
We start with two touching discs, $i$ and $j$, where
one of them, say $i$, remains fixed, while the other is subject
to a force $\vec{F}$ perpendicular to its surface (no external torque 
is induced) along the $x$-axis.
As a result of this external force,
the disc $j$ undergoes rotation.
The contact force is obtained from the corresponding springs that are
computed, as described in Sec.~\ref{sec:model}, 
and acts against the external force.
This results in an oscillation of disc $j$ till 
relaxation (dashed circle in Fig.~\ref{fig4}) with a
final center of mass displacement of $\Delta R$ and a rotation
around the center of mass of $\Delta\theta$.
Since $\vec{F}$ is kept constant, the procedure is stress controlled.

For the two discs sketched in Fig.~\ref{fig4}, we plot in Fig.~\ref{fig5}
the kinetic energy as a function of time, from the beginning until relaxation,
using different integration steps, namely 
$\Delta t=10^{-1}, 2\times 10^{-1}, 10^{-2},
10^{-3}$ and $10^{-4}\hbox{\ s}$ in units of $t_c$. 
As we see, the kinetic energy decays exponentially,
\begin{equation}
E_k(t) = E_k^{(0)}\exp{\left ( -\tfrac{t}{t_R(\Delta t)}\right )} ,
\label{kinen_decay}
\end{equation}
where $t_R$ is a relaxation time whose value clearly depends 
on the integration step $\Delta t$.
As illustrated in the inset of Fig.~\ref{fig5}, 
this change in $t_R$ is not observed when friction is absent
($\mu=0$), since no tangential forces are considered ($F_e^t=0$).
\begin{figure}[htb]
\begin{center}
\includegraphics*[width=8.3cm,angle=0]{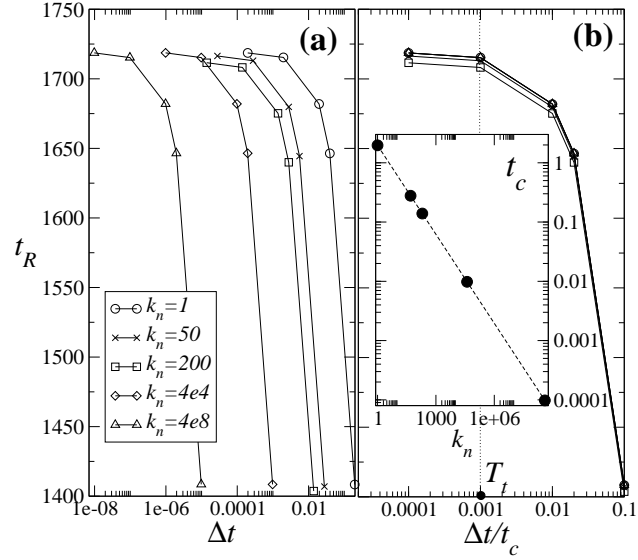}
\end{center}
\caption{\protect 
     The relaxation time $t_R$ (in units of $t_c$)
     as a function of
     {\bf (a)} the integration step $\Delta t$ and
     {\bf (b)} the normalized integration step $\Delta t/t_c$,
     where the contact time $t_c$ is defined in Eq.~(\ref{tc}).
     Here the friction coefficient is kept fixed $\mu=500$
     and different stiffnesses $k_n$ (in units of N/m) are considered.
     The quotient $\Delta t/t_c$ collapses all the curves
     for different $k_n$. We find $t_c\sim k_n^{-1/2}$ as 
     illustrated in the inset (see Eq.~(\ref{tc})).
     As a final result one finds a constant $T_t=10^{-3}$
     (dashed vertical line).
     For other values of the friction coefficient one
     observes similar results.}
\label{fig6}
\end{figure}

Next, we will show that this dependence of $t_R$ on $\Delta t$
vanishes for 
\begin{equation}
\Delta t \lesssim T_t(k_n,\mu) t_c ,
\label{AA}
\end{equation}
where $T_t(k_n,\mu)$ is a specific function that is
determined below.
Notice that, in our case, the only free parameters on which $T_t$ may depend 
are the friction $\mu$ and the normal stiffness $k_n$, since we 
consider a fixed restitution coefficient in the normal direction, 
$\epsilon_n=0.9875$ and fixed relations $k_t/k_n=1/3$ and 
$\epsilon_t/\epsilon_n=1.0053$.

Figure \ref{fig6}a shows the relaxation time $t_R$ of the kinetic
energy of the two-particle system for different values of
stiffnesses, namely for $k_n=1, 50, 200, 10^4$ and $10^8 \hbox{\ N/m}$.
For all $k_n$ values, one sees that, for decreasing $\Delta t$, 
the relaxation time $t_R$ increases until it converges to a maximum.
The stabilization of $t_R$ occurs when $\Delta t$ is small
compared to the natural period $1/\omega_0$ of the system.
We define $T_t$ as the largest value of $\Delta t$ 
for which we have this maximal relaxation time.

As shown in Fig.~\ref{fig6}b, all curves in 
\ref{fig6}a can be collapsed by using the normalized integration
step $\Delta t/t_c$.
From Eq.~(\ref{tc}) we calculate the contact times corresponding to
these $k_n$ values as $t_c=1.969, 0.278, 0.139, 9.8\times 10^{-2}$ 
and $9.8\times 10^{-5}\hbox{\ s}$ respectively. 
In fact, as shown in the inset of Fig.~\ref{fig6}b the relaxation
time scales with the stiffness as $t_c\sim k_n^{-1/2}$ 
(see Eq.~(\ref{tc})).

From Fig.~\ref{fig6} one can conclude that the 
relaxation time converges when the integration step
obeys Eq.~(\ref{AA}) with $T_t=10^{-3}$ (dashed vertical line
in Fig.~\ref{fig6}b). 
We simulate the system also for $\mu=0.005, 0.005, 0.05, 0.5, 5, 50$ 
and $500$ and similar results were obtained. 

In Fig.~\ref{fig6} both translation and rotation of particles 
are considered. 
The rotation of particles is usually of crucial interest, 
for instance to simulate rolling~\cite{fernando06,latham05}.
But, suppressing rotation can also be of interest for instance, when 
simulating fault gouges: by hindering the rotation of 
particles, one can mimic young faults where a strong interlocking 
between the constituent rocks is expected~\cite{fernando06}.

\begin{figure}[H]
\begin{center}
\includegraphics*[width=8.3cm,angle=0]{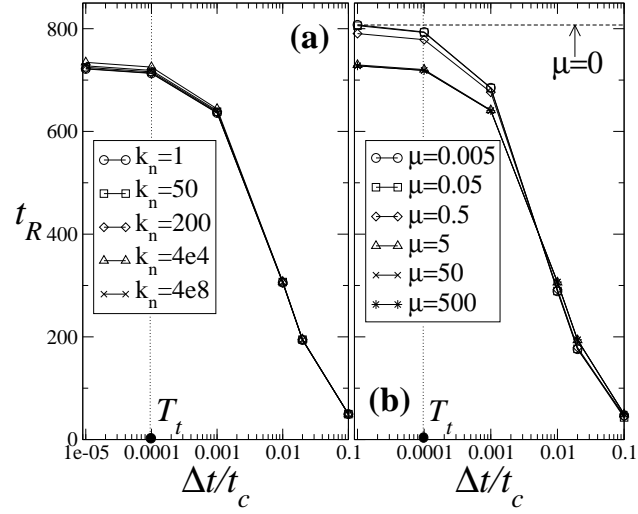}
\end{center}
\caption{\protect 
        The relaxation time $t_R$ (in units of $t_c$) of the kinetic energy
        as a function of the normalized integration step $\Delta t/t_c$,
        when rotation is suppressed.
        {\bf (a)} $\mu=500$ and different values of $k_n$
        and for
        {\bf (b)} $k_n=4\times 10^8$ and different values of
        $\mu$. The dashed horizontal line $\mu=0$ in (b) indicates the 
        relaxation time of the kinetic energy in the absence 
        of friction (see text).}
\label{fig7}
\end{figure}

To study this scenario, we present in Fig.~\ref{fig7}a the relaxation 
time for the same parameter values as in Fig.~\ref{fig6}, now
disabling rotation.
Here, we obtain a constant $T_t=10^{-4}$ also, independent of 
$k_n$, one order of magnitude 
smaller than the previous value in Fig.~\ref{fig6}.
In other words, when rotation is suppressed, one must consider
integration steps typically one order of magnitude smaller than in 
the case when the discs are able to rotate.
This can be explained as follows.

When suppressing rotation, one restricts the system to have
a single degree of freedom. All energy stored in the rotational 
degree of freedom through the integration of the equations of motion
is suppressed. This effectively acts like an increase of the
friction coefficient, 
making the system more sensitive to the integration step, i.e.~yielding
a smaller bound $T_t$. 
By comparing Fig.~\ref{fig7}a with Fig.~\ref{fig6}a, one
sees that the relaxation time $t_R$ is smaller when rotation
is suppressed.


From the bounds on the integration steps obtained
above, one realizes that, in general, the correct integration step
must be significantly smaller than usually assumed. 
\begin{figure}[H]
\begin{center}
\includegraphics*[width=8.3cm,angle=0]{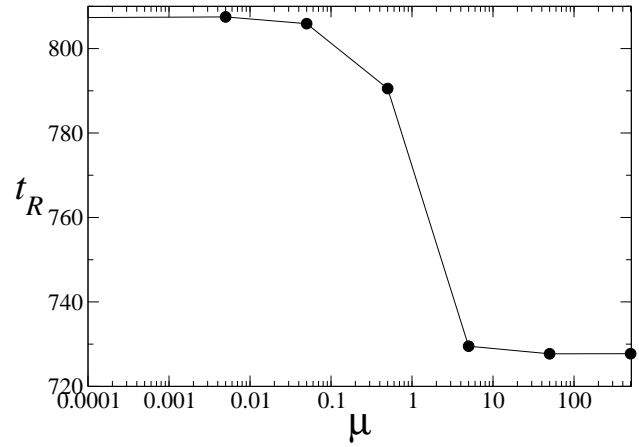}
\end{center}
\caption{\protect 
        The relaxation time $t_R$ (in units of $t_c$) as a 
        function of the friction coefficient $\mu$ when
        rotation is suppressed. Here
        $k_n=4\times 10^8 \hbox{\ N/m}$ which corresponds
        to a contact time $t_c=9.8\times 10^{-5}\hbox{\ s}$.
        The normalized integration step is
        $\Delta t/t_c = 10^{-5}$.}
\label{fig8}
\end{figure}

While Fig.~\ref{fig7}a clearly shows that 
$t_R$ does not depend on the stiffness $k_n$, from Fig.~\ref{fig7}b
one sees that the same is not true for the friction coefficient
$\mu$.
Indeed, from Fig.~\ref{fig8} one sees that there is a change
of the relaxation time around $\mu=1$. 
Here, the values  correspond to a normalized integration step 
$\Delta t/t_c = 10^{-5}$ for which $t_R$ has already converged.
This might be explained by considering the fact that for large values 
of $\mu$ the contact is essentially non-sliding, which induces a faster 
relaxation than for smaller $\mu$ values.

It is important to stress
that all the results above were taken within the elastic regime,
since the dependence on $\Delta t$ does not occur when the Coulomb
condition is fulfilled (inelastic regime).
This fact indicates that the improvements in the algorithm
should be implemented when computing the elastic component of the
tangential contact force, in Eq.~(\ref{elasticelong}),
as explained in the next Section.

\section{Improved approach to integrate the contact force}
\label{sec:geometrical}

In this Section we will describe a technique to overcome 
the need of very small integration steps.
As shown previously, when using Cundall's spring\cite{cundall89},
the relaxation time of the two particles only converges when 
$\Delta t$ is a small fraction $T_t$ of the contact time $t_c$. 
This is due to the fact that
the elastic elongation is assumed to be linear in $\Delta t$,
i.e.~the finite difference scheme in Eq.~(\ref{elasticelong}) is
very low order, $(\Delta t)^2$,
compromising the convergence of the numerical scheme that is of 
order $(\Delta t)^6$.
Therefore, the most plausible way to improve our algorithm is by
choosing a different expression to compute the elastic tangential 
elongation $\xi$ without using Eq.~(\ref{elasticelong}).


We will introduce an expression for $\xi$ that contains only 
the quantities computed in the predictor step. 
In this way we guarantee that $\xi$ has errors of the order 
of $(\Delta t)^6$, instead of 
$(\Delta t)^2$, as it is the case of Eq.~(\ref{elasticelong}).
Let us illustrate our approach on the simple system of two discs
considered in the previous Section (see Fig.~\ref{fig4}).

On one side, if rotation is not allowed, the elastic elongation $\xi$
depends only on the relative position of the two particles.
In this case we substitute Eq.~(\ref{elasticelong}) by the
expression
\begin{equation}
\xi_j^{(tr)}(t+\Delta t) = \xi_j^{(tr)}(t) 
                    + \frac{a_i}{a_i+a_j}
                    (\vec{R}_{ij}^p(t+\Delta t)-\vec{R}_{ij}^p(t))\cdot \hat{t}^c ,
\label{xi_trans}
\end{equation}
where $a_i$ and $a_j$ are the radii of the discs $i$ and $j$ respectively,
$\vec{R}_{ij}$ is the vector joining both centers of mass
and points in the direction $i\to j$ (see Fig.~\ref{fig4}). 
Index $p$ indicates quantities derived from the coordinates 
computed at the predictor step. 

On the other side, if $\vec{R}_{ij}$ is kept constant and only rotation
is possible, particle $j$ will have an elongation $\xi$ that depends
only on its rotation between time $t$ and the predictor step:
\begin{equation}
\xi_j^{(rot)}(t+\Delta t) = \xi_j^{(rot)}(t) + 
                            (\theta_j^p(t+\Delta t)-\theta_j^p(t))a_j ,
\label{xi_rotation}
\end{equation}
where $\theta^p$ and $\theta(t)$ are the angles of some reference 
point on particle $j$ at the predictor step and at time $t$
respectively.

When both translation and rotation
of particle $j$ occur, then the elongation is the superposition of
both contributions, yielding $\xi_j = \xi_j^{(tr)} + \xi_j^{(rot)}$.

\begin{figure}[t]
\begin{center}
\includegraphics*[width=8.3cm,angle=0]{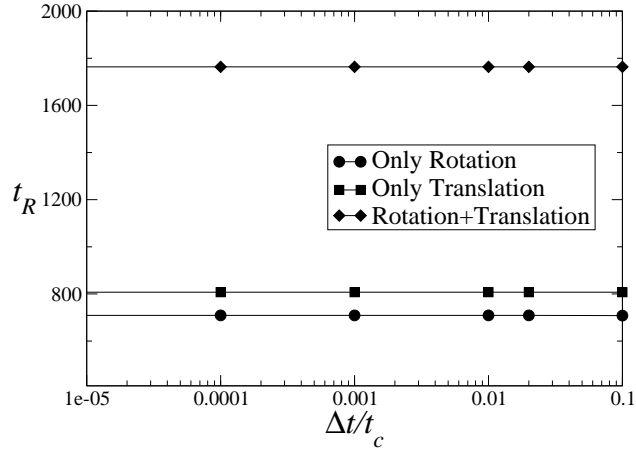}
\end{center}
\caption{\protect 
        The relaxation time $t_R$ (in units of $t_c$) using 
        Eqs.~(\ref{xi_trans}) and (\ref{xi_rotation}) between
        two discs, as illustrated in Fig.~\ref{fig4}.
        For the three cases when considering only rotation, only 
        translation or both, the relaxation time remains constant 
        independent of the integration step.}
\label{fig9}
\end{figure}

Figure~\ref{fig9} shows the relaxation time $t_R$ as a function of
the integration step for the three situations above, namely when
only rotation is considered, when only translation is considered
and when both rotation and translation are allowed.
As we see for all these cases, the relaxation time is independent
on the integration step. 
This is due to the fact that all quantities in the expression for $\xi$
above are computed at the predictor step which has an error of
the order of $(\Delta t)^6$, i.e.~the error $(\Delta t)^2$ introduced 
in Eq.~(\ref{elasticelong}) is now eliminated.
Therefore, with the expressions in Eqs.~(\ref{xi_trans}) and 
(\ref{xi_rotation}) one can use integration steps significantly
larger than with the original Cundall spring.
\begin{figure}[htb]
\begin{center}
\includegraphics*[width=8.3cm,angle=0]{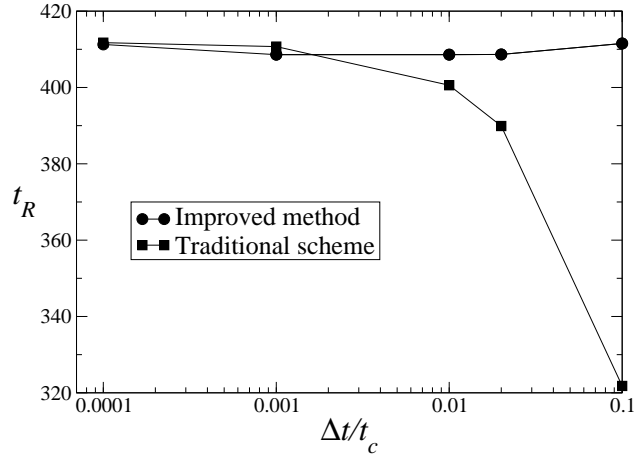}
\end{center}
\caption{\protect 
        Stress control test between two polygonal particles,
        as illustrated in Fig.~\ref{fig1}.
        Comparison of the relaxation time $t_R$ (in units of $t_c$) 
        when using the standard integration scheme (squares)
        and the proposed improved scheme (circles).
        Here, rotation is neglected.}
\label{fig10}
\end{figure}

When considering discs, one does not take into account the shape of the 
particles. Next, we consider the more realistic situation of irregular
polygonal-shaped particles.
Motion of rigid particles with polygonal shape is more complicated
than that of simple discs, since the contact point no longer 
lies on the vector connecting the centers of mass.
Further, for polygons, one must also be careful when
decomposing the dynamics of each particle into translation and rotation 
around its center
of mass. This implies recalculating each time the position of
the center of mass (only from translation) and the relative position
of the vertices (only from rotation).

Therefore, for the translational contribution $\xi^{(tr)}$ in
Eqs.~(\ref{xi_trans}), we compute the overlap area between the two 
particles at $t$ and at the predictor step. The overlap area is in general
a polygon with a geometrical center that can be computed also at
time $t$ and the subsequent predictor step, yielding $\vec{r}_c (t)$ 
and $\vec{r}_c^{p}$ respectively. The increment for the
translational contribution will be just the tangential projection
$(\vec{r}_c (t)-\vec{r}_c^{p})\cdot \hat{t}^c$. 
Similarly, the contribution from the (polygonal) particle rotation
is computed by determining branch vectors, $\vec{r}_b (t)$ and 
$\vec{r}_b^{p}$, defined as the vectors joining the center of the particle
and the center of the overlap area at time step $t$ and the predictor
step respectively.
Computing the branch vector at $t$ and at the predictor step, one derives
the angle defined by them, namely
$\theta=\arccos{\left ( \vec{r}_b^{p}\cdot\vec{r}_b (t)/
                (r_b^{p} r_b (t)) \right )}$ and the
average value $(r_b^p-r_b(t))/2$, yielding
an increment in Eq.~(\ref{xi_rotation}) given by
$\theta (r_b^p-r_b(t))/2$.

Figure \ref{fig10} compares how the relaxation time varies with
the normalized time step when the original Cundall approach is
used (squares) and when our improved approach is introduced
(circles). Clearly, the dependence on the integration step
observed for the usual integration scheme disappears when
our improved approach is introduced.
Therefore, all the conclusions taken above for discs remain 
valid for polygons.

\section{Discussion and conclusions}
\label{sec:conclusions}

In this paper we introduced a technique to improve
the accuracy of the numerical scheme used to compute the evolution 
of particle systems. 

To that end, we have first shown that the range of admissible
integration steps has an upper limit significantly smaller
than typically used.
The accuracy of the numerical scheme not only depends on
the associated error when computing the particle positions
(predictor-corrector scheme),
but also on the accuracy when determining the frictional
force, which is usually implemented by the Cundall spring.
Since the Cundall spring is linear in the
integration step, the overall accuracy of the numerical scheme
cannot be higher than $(\Delta t)^2$.
Therefore, when large integration steps are required,
e.g.~in slow shearing, the numerical scheme does not give
accurate results.

To overcome this problem we introduced an alternative
approach for computing the frictional forces
that suits not only the simple 
situation of discs but the more realistic situation
of polygonal particles.
Our approach is particularly suited for situations where
non-sliding contacts are relevant to the overall response.
In general, for any other integration scheme, the substitution 
of the Cundall spring expression by the relations 
introduced in Eqs.~(\ref{xi_trans})~and~(\ref{xi_rotation}), 
yields an error that is of the same order of 
the one associated with the predictor-corrector scheme.


Inspired by the above results, some questions arise
to further improve  our approach.
First, 
the influence of the relations $k_t/k_n$ and
$\epsilon_t/\epsilon_n$ should also be considered. Preliminary 
simulations have shown that the upper limit for the integration step
increases with the value of $k_t/k_n$.
Second, the test assumes a unique choice for the
position of the contact point.
However, in a system under shearing the integration must be 
also performed before the appearance of new contacts.
The initial contact point of a new contact will
depend on the size of the integration step.
This point should also be taken into account within our new approach, 
either by assuming some sort of interpolation or by using an 
event-driven scheme till the first contact point.
Third, there is the problem of how to better define
the contact point between two polygons.
Since the contact point is taken as the geometrical
center of their overlap area, the 
branch vectors also vary during rotation, which is not
taken into account in our present approach.
These and other points will be addressed in the future.

\section*{Acknowledgments}
The authors thank Fernando Alonso-Marroqu\'{\i}n for useful
discussions.
We thank support by German-Israeli Foundation and
by {\it Deutsche Forschungsgemeinschaft}, under the project
HE 2732781.
PGL thanks support by {\it Deutsche Forschungsgemeinschaft}, under
the project LI 1599/1-1. HJH thanks the Max Planck prize.



\begin{thebibliography}{00}

\bibitem{poeschel05} T.~P\"oschel and T.~Schwager,
                     {\it Computational Granular Dynamics}
		     (Springer, Berlin, 2005).

\bibitem{ciamarra05} M.P.~Ciamarra, A.~Coniglio, M.~Nicodemi,
                     Phys.~Rev.~Lett.~{\bf 94} 188001 (2005).

\bibitem{dacruz05} F.~da Cruz, S.~Eman, M.~Prochnow, J.N. Roux
                     Phys.~Rev.~E.~{\bf 72} 021309 (2005).


\bibitem{cundall89} P.A.~Cundall,
                    Ingenieur-Archiv {\bf 59}, 148 (1989). 

\bibitem{thompson91} P.A.~Thompson, G.S.~Grest,
                     Phys.~Rev.~Lett.~{\bf 67} 1751 (1991).

\bibitem{pena07} A.A.~Pe\~na, R.~Garc\'{\i}a-Rojo, H.J.~Herrmann,
                 Granular Matter, ~{\bf 9}:279-291 (2007).

\bibitem{fernando06} F.~Alonso-Marroqu\'{\i}n, I.~Vardoulakis,
                     H.J.~Herrmann, D.~Weatherley, P.~Mora,
                     Phys.~Rev.~E {\bf 74}, 031306 (2006).


\bibitem{mora99} P.~Mora, D.~Place,
                 Geophys.~Res.~Lett.~{\bf 26}, 123 (1999).



\bibitem{mcnamara07} S.~McNamara, R.~Gar\'{\i}a-Rojo and
                 H.J.~Herrmann, submitted to Phys.~Rev.~E, 2007.

\bibitem{allen03} M.P.~Allen and D.J.~Tildesley,
                  {\it Computer Simulation of Liquids}
                  (Oxford Univ.~Press, Oxford, 2003).

\bibitem{luding94} S.~Luding, E.~Cl\'ement, A.~Blumen, J.~Rajchenbach, 
                   J.~Duran,
                   Phys.~Rev.~E {\bf 50}, 4113 (1994). 

\bibitem{cundall79} P.A.~Cundall and O.D.L.~Strack,
                    G\'eotechnique {\bf 29}, 47-65 (1979).

\bibitem{rougier04} E.~Rougier, A.~Munjiza, N.W.M.~John,
                    Int.~J.~Numer.~Meth.~Eng.~{\bf 61}, 856 
                    (2004). 

\bibitem{foerster94} S.F. ~Foerster, M.Y.~Louge, H.~Chang, K. Allia,
                     Phys. Fluids {\bf 6(3)}, 1108 (1994).

\bibitem{luding98} S.~Luding,
                    in 
                    {\it Physics of Dry Granular Media}, 
                    H.J.~Herrmann, J.-P.~Hovi, and S.~Luding
                    (Kluwer Academic Publishers, Dordrecht, 1998),
                    pp.~285.

\bibitem{matuttis98} H.-G.~Matuttis,
                     Granular Matter {\bf 1}, 83 (1998).

\bibitem{tillemans95} H.J. Tillemans, H.J Herrmann,
                     Physica A {\bf 217}, :\penalty0 261-288 (1995).

\bibitem{alonso04} F.~Alonso-Marroqu\'{\i}n and H.J.~Herrmann,
                   Phys.~Rev.~Lett.~{\bf 92}, 054301 (2004).

\bibitem{latham05} S.~Latham, S.~Abe, P.~Mora,
                   in R.~Garc\'{\i}a-Rojo, H.J.~Herrmann, and 
                   S.~McNamara (eds.),
                   {\it Powders and Grains 2005}, 
                   (Balkema, Stuttgart, 2005) pp.~213.



\end{thebibliography}
\end{document}